\documentclass[final,5p,times,twocolumn]{elsarticle}
\usepackage{amsmath,amssymb}
\usepackage{graphicx}
\usepackage{color}
\usepackage{bm}
\usepackage{txfonts}
\usepackage{xspace}  
\usepackage{multirow}
\usepackage{dcolumn} %centers to the decimal point in tables using ``d''
\usepackage{xspace}  
\newcommand{\etal}{\textit{et al.}\xspace}

\journal{Physics Letters B}

\begin{document}

%\preprint{}

\title{Triaxiality and Vibrational Modes in Strongly Deformed Nuclei} % no punctuation

\author[cns,tsuk]{Y.~Tsunoda}
\ead{ytsunoda@nucl.ph.tsukuba.ac.jp}
\author[ut,riken,gan,tud]{T.~Otsuka}
\ead{otsuka@phys.s.u-tokyo.ac.jp}
\author[tsuk]{N.~Shimizu}
\author[sac]{T.~Duguet}
\author[jar,cns]{Y.~Utsuno}
\author[ke]{T.~Abe}

\address[cns]{Center for Nuclear Study, The University of Tokyo, 7-3-1 Hongo, Bunkyo, Tokyo 113-0033, Japan}
\address[tsuk]{Center for Computational Sciences, University of Tsukuba, 1-1-1 Tennodai, Tsukuba, Ibaraki, 305-8577, Japan}
\address[ut]{Department of Physics, The University of Tokyo, 7-3-1 Hongo, Bunkyo, Tokyo 113-0033, Japan}
\address[riken]{RIKEN Nishina Center, 2-1 Hirosawa, Wako, Saitama 351-0198, Japan}
\address[gan]{Grand Acc\'el\'erateur National d’Ions Lourds, CEA/DRF-CNRS/IN2P3, Bvd Henri Becquerel, F-14076 Caen, France}
\address[tud]{Institut f\"ur Kernphysik, Technische Universit\"at Darmstadt, D-64289 Darmstadt, Germany}
\address[sac]{IRFU, CEA, Université Paris-Saclay, 91191, Gif-sur-Yvette, France} 
\address[jar]{Advanced Science Research Center, Japan Atomic Energy Agency, Tokai, Ibaraki 319-1195, Japan} 
\address[ke]{Quantum Computing Center, Keio University, 3-14-1 Hiyoshi, Kohoku-ku, Yokohama, Kanagawa 223-8522, Japan} 

\date{\today}

\begin{abstract}   
Low-energy vibrational excitations associated with the fluctuation of quadrupole deformed shapes are discussed within the framework of state-of-the-art Configuration Interaction calculations, actually performed via the Quasi-particle Vacua Shell Model version of the Monte Carlo Shell Model. Recently, low-lying $\gamma$ bands in heavy strongly deformed nuclei were shown to be rotational $K^P$ = 2$^+$ excitations of triaxially deformed states (see T. Otsuka {\it et al.}, Eur. Phys. J. A 61, 126 (2025)) rather than vibrational excitations as traditionally interpreted. \textcolor{black}{We present additional illustrations as to how the E2 excitation strength to the gamma band changes as the triaxial deformation parameter $\gamma$ increases, as a systematic indicator for the evolving triaxiality. This is a part of a general formulation of E2 properties of strongly-deformed triaxial states obtained within the quantum many-body scheme. Once the triaxial picture is clarified as a whole,} it is important to identify possible low-lying \textcolor{black}{$\gamma$ and other} vibrational excitations and to characterize the excitation energy at which they emerge. Focusing on two typical examples, $^{166}$Er and $^{162}$Dy, vibrational states are indeed identified above the $\gamma$ band using an extended version of the so-called T-plot. \textcolor{black}{Notable agreement with experiment is seen for the $\gamma$-vibrational band in $^{166}$Er and for the $\beta$-vibrational band in $^{162}$Dy, regarding energy levels and $E2$ transitions.} The phenomenon of shape coexistence is also shown to produce low-lying states below such vibrational band heads. These results suggest novel and rich structures in heavy deformed nuclei, \textcolor{black}{opening doors to future dedicated studies.} %While experimental counterparts are seen for some of such states, others are predictions opening doors to future dedicated experiments. 
\end{abstract}

\maketitle

\section{Introduction}
Low-energy rotational excitations emerge in nuclei characterized by strongly deformed ellipsoidal shapes, as pointed out early on by Rainwater \cite{rainwater1950}, as well as by Bohr and Mottelson \cite{bohr1952,bohr_mottelson1953,aage_bohr_nobel}.  Properties associated with strong deformation were discussed in many works afterwards as described in textbooks, e.g. see Refs.~\cite{bohr_mottelson_book2,rowe_book,deshalit_book,preston_bhaduri_book,ring_schuck_book,eisenberg_greiner_book1,casten_book}.      
Complementarily, low-energy vibrational excitations arise in various situations, including in nuclei characterized by deformed ellipsoidal shapes. Well-known examples are $\beta$- and $\gamma$-vibrations \cite{aage_bohr_nobel,bohr_mottelson_book2}.  There have been many extensive studies of vibrational modes, as also documented in the textbooks mentioned above.

Recently, a systematic investigation of strongly deformed quadrupole shapes and associated rotational excitations was performed in heavy ($A=Z+N>140$) nuclei with even neutron ($N$) and proton ($Z$) numbers~\cite{epja,otsuka_2025A}, after an early attempt~\cite{otsuka_2019}. This was achieved using state-of-the-art ultra-large-scale Configuration Interaction (CI) calculations performed within the frame of the Quasi-particle Vacua Shell Model (QVSM)~\cite{shimizu_2021} version of the Monte Carlo Shell Model (MCSM) \cite{mcsm2001,shimizu2012}. In such a study, the band head of so-called $\gamma$ bands, which is traditionally interpreted as %triaxial  
$\gamma$ vibration, has been convincingly shown to rather be the $K^P$ = 2$^+$ rotational excitation of a triaxially deformed ground state. \textcolor{black}{We present a new illustration on this point based on the behavior of E2 transition between the ground states and these $K^P$ = 2$^+$ excitations over many nuclei.  This illustration is obtained within a quantum many-body framework, although it is inspired from the model by Davydov {\it et al.} \cite{Davydov1958}. } 

This change of perspective poses the question, addressed in the present Letter, of the emergence of low-lying vibrational excitations in such heavy strongly deformed nuclei. To do so, the MCSM calculation is extended  
with additional MCSM basis vectors compared to earlier studies focusing on rotational states, with the goal to target vibrational states of quadrupole character at excitation energies below $\sim$ 4 MeV. The goal here is to uncover novel features exhibited by vibrational excitations \cite{bohr_mottelson_book2,rowe_book,deshalit_book,preston_bhaduri_book,ring_schuck_book,eisenberg_greiner_book1,casten_book,frauendorf_2015} and shape-coexistence phenomena \cite{wood1992,andreyev2000,heyde2011,leoni2024}. To do so, $^{166}$Er and $^{162}$Dy are presently used as typical illustrative examples.

%%%  CI simulation of  nuclear shape   %%%
\section{CI calculation}
The MCSM~\cite{mcsm2001,shimizu2012} solves the many-body Schr\"odinger equation within a designated %restricted 
many-body Hilbert space built on top of a frozen core based on a valence single-particle %one-nucleon 
space in which a given number of valence nucleons interact via an effective two-nucleon (2N) interaction. The valence proton (neutron) space used in the present calculation collects single-particle states belonging to the $sdg$($pfh$) harmonic-oscillator (HO) shell plus the lower half of the next shell, i.e. it is made of the one-and-a-half HO shell on top of a $^{110}$Zr inert core. The dimension of the many-body Hilbert space is about 
10$^{33}$, which is formidably larger than the current limit ($\sim$ 10$^{11}$) of conventional CI codes. The proton-proton and neutron-neutron channels of the 2N interaction are taken from Ref.~\cite{brownPb} and completed by the V$_{\mathrm{MU}}$ interaction~\cite{otsuka2010} for matrix elements not covered by this recipe. The V$_{\mathrm{MU}}$ interaction, also used for the proton-neutron channel (including a slight modification), was determined as a simple modeling~\cite{otsuka2010} for systematic studies of Zr, Sn, Nd, Sm and Hg isotopes
\cite{togashi2016,kremer2016,togashi2018,marsh2018,sels2019,otsuka_2019,tsunoda_2023}.  
The QVSM~\cite{shimizu_2021} version of the MCSM~\cite{mcsm2001,shimizu2012} utilizes Bogoliubov basis vectors so that pairing correlations can be incorporated within individual basis vectors, as this is a serious problem for heavy nuclei with Slater determinants as in the usual MCSM. 

For a given $(J, M)$ value, where $J$ ($M$) denotes the total angular momentum (its $z$-component), the $k$-th eigenstate reads in the QVSM as
\begin{equation}
 |\Psi_k\rangle\equiv\sum_{iK} f^{(k)}_{iK} |\phi_{iK}\rangle \, ,    %{\mathcal P}_{J,M,K} |\phi\rangle
%\Psi \, = \, \sum_i \, f_i \, \hat{{\mathcal P}}_{J^{P}} \, \phi_i \,\,, 
\label{eq:mcsm_psi}
\end{equation}
with $f^{(k)}_{iK}$ being the amplitudes of the mixing and 
\begin{equation}
|\phi_{iK}\rangle \equiv  {\mathcal P}^J_{MK} |\phi\rangle \, 
\label{eq:mcsm_psi2}
\end{equation}
denoting the $i$-th MCSM basis vector. The latter is obtained by projecting the triaxially deformed number-projected Bogoliubov state $|\phi\rangle$ onto good $J, M$ (parity projection is also included, but not explicitly shown for brevity) via the symmetry projection operator ${\mathcal P}^J_{MK}$. The integer $K$ runs from $-J$ to $J$ and thus corresponds to independent basis vectors for a given set of $J$ and $M$ values.

Although the geometrical meaning of $K$ quantum number may be lost in general MCSM calculations via the mixing of the $K$ values, the meaning can in fact be retained to good approximation \cite{epja}. 
\textcolor{black}{For strongly quadrupole deformed cases, each basis vector is handled as the intrinsic state with the ellipsoidal shape specified by quadrupole moments of this basis vector.   
The $z$-axis, i.e., the longest ellipsoidal axis, is aligned to a common direction among the basis vectors, without changing MCSM results.  
The $K$ quantum number defined with this $z$-axis carries a physical meaning like that in geometrical models.  
%Note that more correlations are included than such simpler models.  
The $K$ quantum number thus defined} was shown to be practically conserved in low-lying strongly deformed states, even with triaxiality \cite{epja}.  This feature is confirmed to similarly remain in vibrational states to be discussed for which $K$ quantum numbers are indicated. 
%These $K$ quantum number exhibits similar physical meanings to the models/theories of collective bands.  

%%%%%%%%%%%%%%%%%%%%%%%%%%%%%%%

% Fig1: 166Er
\begin{figure}[bt]
%  \centering
  \includegraphics[width=9.0cm]{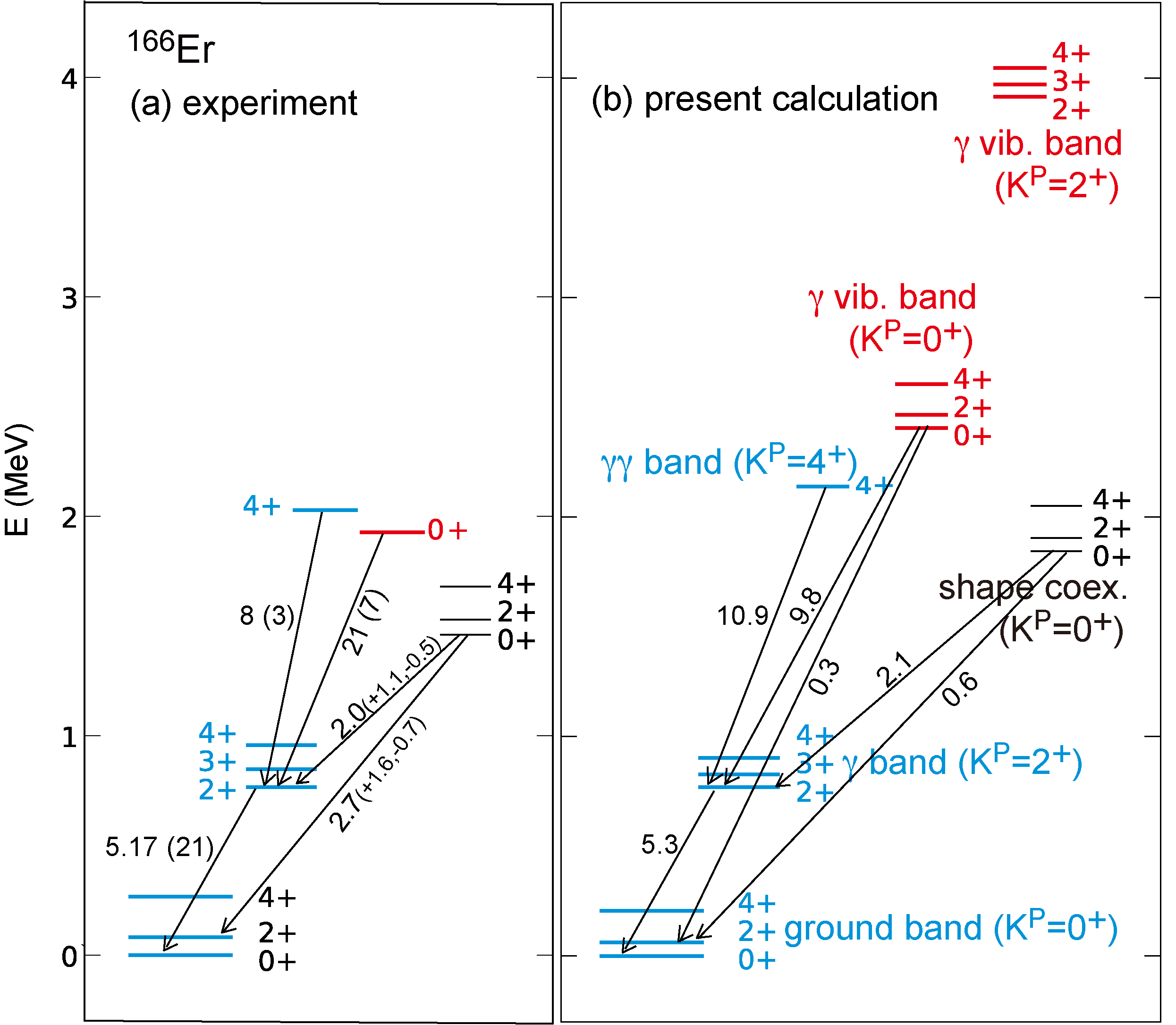}  
  \caption{Experimental~\cite{ensdf,garrett_1997} (a) and present theoretical (b) energy levels in $^{166}$Er. 
  Shown in blue are rotational states of $K^P$=0$^+$, 2$^+$, 4$^+$ obtained by the present theory, and are the corresponding experimental ones.  
  Bands built on vibrational excitation modes (see Sec.~\ref{PT_166Er}) are shown in red 
  including an experimental counterpart.  The other is shown in black.
  Many experimental levels above $\sim$2 MeV are not shown for clarity, and some high-lying theoretical levels are also omitted because of their weaker relevance.  Selected $B(E2)$ values are shown in W.u.\cite{ensdf,garrett_1997,nudat3} Quite a few transitions and quadrupole moments are omitted, because they were presented in \cite{epja}. 
  }   
  \label{fig:level_166Er}  
\end{figure}  

%%%%%%%%%%%%%%%%%%%%%%%%%%%%%%%

%%%%%%%%%%%%%%%%%%%%%%%%%%%%%%%

% Fig2: B(E2) to 2+gamma
\begin{figure}[bt]
  \centering
  \includegraphics[width=8.6cm]{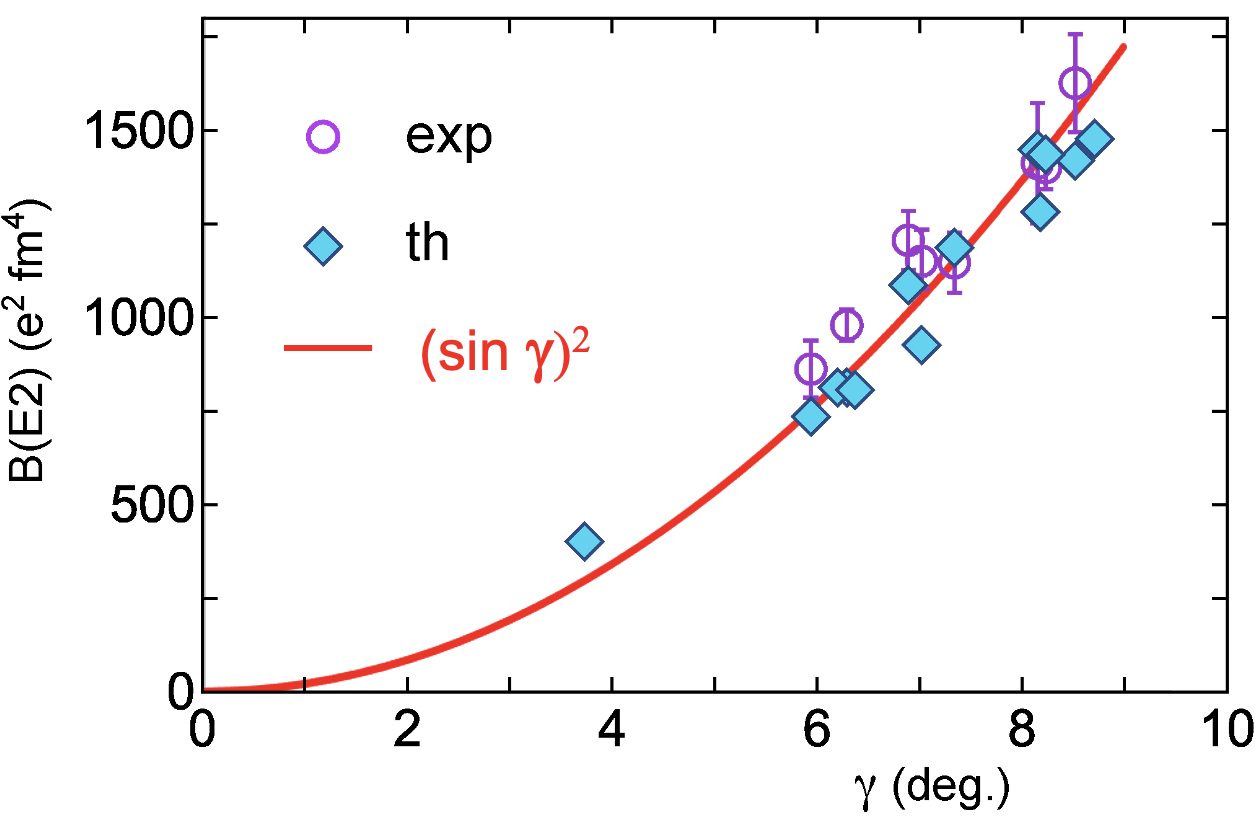}  %166Er_levels_9red
  \caption{
  \textcolor{black}{Systematics of experimental  (open circle)~\cite{ensdf} and theoretical (blue
diamond) B(E2; 0$^+_1$$\rightarrow$2$^+_{\gamma}$) values in e$^2$fm$^4$.  
Both plots are made as a function of the deformation parameter $\gamma$ of the nuclei, from right to left, 
164Yb ($\gamma$=8.7$^{\circ}$), 162Er (8.5$^{\circ}$), 166Er (8.2$^{\circ}$), 168Hf (8.2$^{\circ}$),
164Er (8.2$^{\circ}$), 164Dy (7.3$^{\circ}$), 160Dy (7.0$^{\circ}$), 162Dy (6.9$^{\circ}$), 166Yb (6.4$^{\circ}$), 160Gd (6.3$^{\circ}$), 162Gd (6.2$^{\circ}$), 158Gd (5.9$^{\circ}$), 154Sm (3.7$^{\circ}$).
The $\gamma$ values are for the ground state.  They are obtained by the present theoretical calculation, and are  screened by the criterion that a reasonable agreement is obtained between experiment and theory for individual nucleus. 
The red curve indicates the values obtained by $({\rm sin} \, \gamma)^2$ as it is derived in a microscopic quantum many-body framework (see the text).  
}
Taken and modified from \cite{epja} with kind permission of European Physical Journal.}   
  \label{fig:BE2_gamma}  
\end{figure}  

%%%%%%%%%%%%%%%%%%%%%%%%%%%%%%%

%%%%%%%%%%%%%%%%%%%%%%%%%%%%%%%

\section{Triaxiality in heavy deformed nuclei \label{sec:triaxial}}

We next discuss the shapes and structures of heavy deformed nuclei, which are classified through 
the ratio R42=Ex(4$^+_1$)/Ex(2$^+_1$) $\ge$ 3.00 as in \cite{epja}.

Figure~\ref{fig:level_166Er} displays, for $^{166}$Er taken as an example, experimental \cite{ensdf} and calculated excitation energies of low-lying positive-parity states with angular momenta $J$ = 0, 2, 3, 4 \footnote{States carrying larger $J$ values are not shown for clarity.}. Selected $B(E2)$ reduced transition probabilities among some of these states are also shown. Levels below $\sim$ 1 MeV were discussed in \cite{epja} along with the experimental 4$^+$ state at excitation energy $E_x$ = 2.028 MeV and the calculated $J^P$ = $K^P$ = 4$^+$ state. 

Some of these low-lying states were analyzed in the past from the viewpoint of possible triaxial shapes or so-called $\gamma$-instability, see Refs.~\cite{Davydov1958,Davydov1959,cline_1986,kotlinski_1990,fahlander_1992,sun_2000,sun_2002,delaroche_2010,li_2010,Sharpey-Schafer2019,yang_2021,frauendorf_2024,frauendorf_2025}, as well as to search for double $\gamma$-phonon ($\gamma\gamma$) $K^P$ = 4$^+$ and $K^P$ = 0$^+$ states in~\cite{fahlander_1996,garrett_1997,garrett_2018}. On the other hand, no triaxiality was suggested for strongly-deformed heavy nuclei in, for instance, Refs.~\cite{bohr_mottelson_book2,moller_1995,moller_2006}. A key point of Ref.~\cite{epja} was to demonstrate that the band head of the $K^P$ = 2$^+$ $\gamma$ band, denoted by 2$^+_{\gamma}$ hereafter, relates to the rotation of a triaxial state contrary to the traditional interpretation at play. 

We then look into the prevailing triaxiality in heavy deformed nuclei, because it is one of the essential grounds for formulating vibrational excitations on top of triaxial states.   As indicated in \cite{epja}, the substantial triaxiality is seen in virtually all strongly deformed nuclei, including $^{166}$Er, due to the rotational-symmetry restoration (particularly with the K quantum number) and the role played by specific channels of the two-nucleon ($NN$) interaction in nuclei, such as the hexadecupole forces and the monopole part of the tensor forces.  This prevailing triaxiality was examined from the viewpoint of $B(E2; 0^+_1 \rightarrow 2^+_{\gamma})$ values for 13 isotopes of Sm, Gd, Dy, Er, Yb and Hf nuclei in Fig. 18 of \cite{epja}, where the $\gamma$ values were obtained by the standard T-plot analysis of the QVSM eigenstates\footnote{An advanced version of the T-plot analysis is presented later for the description of vibrational excitations.}. Figure~\ref{fig:BE2_gamma} includes the same plot as Fig. 18 of \cite{epja}: a general trend is obvious, and QVSM results agree well with experimental data of 8 nuclei. Although Fig. 18 of \cite{epja} is very indicative of a correlation between $B(E2)$ and $\gamma$ values, 
we now go beyond by unearthing the origin and the implications of this correlation.  

The $E2$ transitions between states generated from a common triaxial intrinsic state, $\phi$ can be discussed, in a fashion somewhat related  to the prescription of Davydov and Filippov \cite{Davydov1958}, while remaining within a full quantum many-body formulation (i.e., without using the Bohr Hamiltonian).  The quadrupole operator is defined, apart from charges, as
\begin{equation}
Q_{2,\mu} = \sqrt{\frac{4\pi}{5}} \,  \sum_{k}  r_k^2 \, Y_{2\mu}(\theta_k , \eta_k),
\label{eq:Q}
\end{equation}
where $k$ denotes the index of all active nucleons, and $\theta_k$ and $\eta_k$ stand for polar coordinates.

The state $|J, M, K \rangle$  with the standard notation of $J, M, K$ is generated from $| \phi \rangle$ as
\begin{equation}
|J, M, K \rangle \,=\, \frac{2J+1}{8\pi^2} \, \int d\bm{\omega} \, \{D^J_{MK}(\bm{\omega})\}^* \, {\hat R}(\bm{\omega}) \, | \phi \rangle,
\label{eq:JMK}
\end{equation}
where $D^J_{MK}$ stands for a Wigner function, 
$\omega\equiv(\alpha, \beta, \gamma)$ denoting the three Euler angles, and 
${\hat R}$($\alpha, \beta, \gamma$) the rotation operator (see \ref{ApA} for details). The ket state in eq.~(\ref{eq:JMK}) is to be properly normalized.
%%%%%
We assume that the intrinsic state $| \phi \rangle$ is so strongly deformed that a small change of its angular direction makes its overlap with $| \phi \rangle$ decreases rapidly. 
Here, the involvement of many particles, which is natural for collective states, is an important factor, as it generally reduces this overlap by the power of the particle number.  In the extreme limit, often denoted as the needle approximation, $\langle \phi \,|\, {\hat R}(\bm{\zeta}) \phi \rangle = \delta(\bm{\zeta})$ is obtained \footnote{To be more precise, because of the $\pi$-rotation symmetry (see e.g. eqs.(38,50) of \cite{epja}), there are other terms but their effects are implicitly included in the final outcome in the following formulation}.  
        
Matrix elements for $Q_{2,\mu}$ can then be written as 
\begin{eqnarray}
\lefteqn{\langle J', M', K' | \,Q_{2,\mu} \,|J, M, K \rangle} \nonumber \\
& = & \frac{(2J+1)(2J'+1)}{(8\pi^2)^2} \,\Bigl\{ \int d\bm{\omega'}  \langle \, {\hat R}(\bm{\omega'})\phi | D^{J'}_{M'K'}(\bm{\omega'}) \Bigr\}  \nonumber \\
&&  \,\,\,\,\, Q_{2,\mu} \, \Bigl\{ \int d\bm{\omega}  [D^J_{MK} (\bm{\omega})]^*  \, |\, {\hat R}(\bm{\omega}) \, \phi \rangle \Bigr\}  ,  \nonumber \\ 
& = & \frac{(2J+1)(2J'+1)}{(8\pi^2)^2} \, \int d\bm{\omega} \nonumber \\
&&   \,\, D^{J'}_{M'K'}(\bm{\omega}) \,[D^J_{MK}  (\bm{\omega})]^* \, \langle \phi | \,{\hat R}^{-1}(\bm{\omega}) \, Q_{2,\mu}  \, {\hat R}(\bm{\omega}) \, | \phi \rangle, \nonumber \\ 
\label{eq:Qme}
\end{eqnarray}
where on the right-hand side, the integrand is assumed to be strongly peaked according to $\langle \phi \,|\, Q_{2,\mu}\,| \,{\hat R}(\bm{\zeta}) \, \phi \rangle = \delta(\bm{\zeta}) \,
\langle \phi \,|\, Q_{2,\mu} \, |\, \phi \rangle$ due to the strong deformation of a system comprising many particles, in line with the needle approximation discussed above for the overlap
(i.e., only $\bm{\omega}$=$\bm{\omega}'$ contributes). 
The following discussions are made at this limit, as it simulates the fast damping of the overlap and other related quantities as $\omega$ moves away from zero.  
%%%%  REMOVED  %%%
%\textcolor{black}{Although $Q$ can change one of single-particle states, the action of $Q$ operator on a state belonging to the Hilbert subspace spanned by the states generated by angular momentum projection of $| \phi \rangle$ is expected to primarily remain within this subspace.   On the other side, the deviations from this strength closure should arise also in the strengths of vibrational excitations, the magnitude of which may be of interest, as discussed in subsequent sections. 

We rewrite the following term in eq.~(\ref{eq:Qme}) as
\begin{equation}
{\hat R}^{-1}(\bm{\omega}) \, Q_{2,\mu}  \, {\hat R}(\bm{\omega}) \,=\, 
 \sum_{\nu} \{D^2_{\mu, \nu}(\bm{\omega})\}^* \, Q_{2,\nu} . 
\label{eq:RQR}
\end{equation}
We can now take the expectation value $\langle \phi | Q_{2\nu} | \phi \rangle$ for $Q_{2\nu}$ in eq.~(\ref{eq:RQR}). When expressing them, three coordinate axes are set to the principal axes of the ellipsoid as usual (see \cite{epja} if needed), and those values are represented in terms of parameters $a_\nu$ in the usual way (i.e., apart from a common scaling factor, $t$, mentioned below).  Taking into account $a_1=a_{-1}=0$, the $a_\nu$'s are further expressed as
\begin{equation}
a_0 =  \, \beta_2 \,{\rm cos} \, \gamma, \,\,\,\,\,\, {\rm and} \,\,\,\,\,\, a_2=a_{-2}= \, \beta_2 \, {\rm sin} \gamma / \sqrt{2},
\label{eq:a02}
\end{equation}
where $\beta_2$ and $\gamma$ stand for the (usual quadrupole) deformation parameters.   

The quadrupole matrix element in eq.~(\ref{eq:Qme}) can now be expressed by   
\begin{eqnarray}
\lefteqn{\langle \phi | \,{\hat R}^{-1}(\bm{\omega}) \, Q_{2,\mu}  \, {\hat R}(\bm{\omega}) \, | \phi \rangle   
}  \nonumber \\
& = & t \beta_2 \, \Bigr\{ \, [D^{2}_{\mu 0}(\bm{\omega})]^* \, {\rm cos} \, \gamma  \nonumber \\
& & \,\,\,\,\,\,\,\,  + \frac{1}{\sqrt{2}} ([D^{2}_{\mu 2}(\bm{\omega})]^* + [D^{2}_{\mu -2}(\bm{\omega})]^*) \, {\rm sin} \, \gamma \, \Bigl\}, 
\label{eq:a03}
\end{eqnarray}
where $t$ is a parameter fixed, e.g., with $\bm{\omega}$=0, so that the overall magnitude of $Q_{2,\mu}$ is properly treated.\footnote{The $t\beta_2$ factor represents all relevant effects including mathematical factors, mean effects of $r^2$ dependence and possible effective charges in $\langle \phi | Q_{2,\mu} | \phi \rangle$, and is independent of $\mu$. } 
Equation~(\ref{eq:Qme}) is then rewritten as,
\begin{eqnarray}
\lefteqn{\langle J', M', K' | \,Q_{2,\mu} \,|J, M, K \rangle } \nonumber \\
\lefteqn{= \, \frac{(2J+1)(2J'+1)}{(8\pi^2)^2}  
 \int d\bm{\omega} \, D^{J'}_{M'K'}(\bm{\omega}) \,[D^J_{MK}(\bm{\omega})]^* \,\,\,\,\,} \nonumber \\
&t \beta_2 \, \Bigr\{[D^{2}_{\mu 0}(\bm{\omega})]^* \, {\rm cos} \, \gamma + \frac{1}{\sqrt{2}} ([D^{2}_{\mu 2}(\bm{\omega})]^* + [D^{2}_{\mu -2}(\bm{\omega})]^*) \, {\rm sin} \, \gamma\Bigr\} .  \nonumber \\  
\label{eq:Qme3}
\end{eqnarray}

The $\pi$-rotation symmetry, $|\phi \rangle$=$\hat{R}(0,\pi,0)|\phi \rangle$ (see, for instance, Eqs.~(38,50) of \cite{epja}),  is assumed as we are discussing ellipsoidal deformations\footnote{The state of the $\pi$-rotation symmetry can be generated from a state without it (see eq.(38) and related text of \cite{epja}).}.  If $K$$\ne$0, $|J, M, K \rangle$ is associated with $|J, M, -K \rangle$ to form the state
\begin{equation}
\overline{| J, M, K \rangle} =  \frac{1}{\sqrt{2}} \, \Bigr\{|J, M, K \rangle + (-1)^{J+K} |J, M, -K \rangle \Bigr\}.   
\label{eq:K-K}
\end{equation}
Hereafter, we assume $K$$\ge$0 and $\overline{|J, M, K=0\rangle}\equiv|J, M, K=0\rangle$.  
Because of the appearance of $K$ and $-K$ on the right-hand-side of Eq.~(\ref{eq:K-K}),  we obtain,
\begin{eqnarray}
\lefteqn{\overline{\langle J', M', K'=2 |} \,Q_{2,\mu} \, \overline{|J, M, K=0 \rangle} }  \nonumber \\ 
\lefteqn{\,= t \beta_2 \frac{(2J+1)(2J'+1)}{(8\pi^2)^2}  \, \int d\bm{\omega} } \nonumber  \\
\lefteqn{\,\,\,\,\,\,\,\,\, {\times} \frac{1}{2} \Bigr\{ D^{J'}_{M' 2}(\bm{\omega}) [D^{J}_{M 0}(\bm{\omega}) D^{2}_{\mu 2}(\bm{\omega})]^* }  \nonumber \\
 & \,\,\,\,\,\,\,\,\, + D^{J'}_{M' -2}(\bm{\omega}) 
 [D^{J}_{M 0}(\bm{\omega}) D^{2}_{\mu -2}(\bm{\omega})]^* \Bigr\} \, {\rm sin} \, \gamma ,\nonumber \\ 
\lefteqn{= t \beta_2 \frac{2J+1}{8\pi^2} (J M 2 M'-M | J' M' ) (J 0 2 2 | J' 2 ) \, {\rm sin} \, \gamma ,} 
\label{eq:Qme5}
\end{eqnarray}
where Eqs.~(\ref{eq:DDint}) and (\ref{eq:DDproduct}) are used.
This implies the reduced matrix element,
\begin{eqnarray}
&\overline{\langle J', K'=2 ||} \,Q_{2} \, \overline{||J, K=0 \rangle} \, \frac{8\pi^2}{\sqrt{(2J+1)(2J'+1)}} \nonumber \\
&= t \beta_2 \sqrt{2J+1} (J 0 2 2 | J' 2 ) \, {\rm sin} \, \gamma ,
\label{eq:reduced1}
\end{eqnarray}
where the normalization constant and the contributions from the $\pi$-rotated Euler angle to relevant integrals are incorporated.
Equation~(\ref{eq:reduced1}) leads, for $J$=0 and $J'$=2, to
\begin{equation}
B(E2; 0^+_1 \rightarrow 2^+, K=2) =  \, |t|^2 \beta_2^2  \, {\rm sin}^2 \, \gamma .
\label{eq:BE2_2gamma}
\end{equation}
Figure~\ref{fig:BE2_gamma} includes a curve given by this formula, assuming $t\beta_2$ to be constant for the displayed nuclei.  In the figure, its value is adjusted to $B(E2)$ =1750 e$^2$fm$^4$ for $\gamma$=9$^{\circ}$. In addition to matching the experimental/theoretical value for $^{166}$Er, the law given by Eq.~(\ref{eq:BE2_2gamma}) reproduces well the trend. 
Individual deviations in theoretical results are due to many-body correlations beyond the pure rotational picture accounted for by the QVSM calculation.

The parameter $t \beta_2$ is indeed expected to be almost constant among the strongly deformed nuclei under consideration even if a slight trend with $A$ may exist. Thus, the evolution of these $B(E2)$ values seem to be dominated by the $\gamma$ dependence according to the sin$^2 \gamma$ function, i.e. the triaxiality manifests the key factor 
determining the physics at play.   
Although the same $\gamma$ value is assumed between the bra and ket states in the above argument, $\gamma$ may differ between the 0$^+_1$ and 2$^+_{\gamma}$ states. The former is taken in Fig.~\ref{fig:BE2_gamma}. Such difference seems to be smeared out in the E2 matrix elements.  

A formula similar to Eq.~(\ref{eq:RQR}) is put forward in the work by  Davydov and Filippov \cite{Davydov1958} \footnote{The equation differs by a complex conjugate that can be eliminated by using the complex conjugate wave function; see Eq.(~A.21) of \cite{ring_schuck_book}.}. Results are however obtained based a Hamiltonian corresponding to the quantization of a freely rotating classical triaxial rigid-body, that produces energy levels, e.g., 4$^+_{\gamma}$ (see Fig. 1 of \cite{Davydov1958}), differing significantly from experiment and from those produced in the present work\footnote{The formulas for E2 transitions in \cite{Davydov1958} includes $K$-mixings by this Hamiltonian, and look different from the corresponding ones in the present work, where the quantum many-body system is treated with conserved $K$ quantum number.}. Thus, the connection between the present treatment of E2 transitions and that in the Davydov model exists but is limited. 

The present calculation can be extended to the decay of the $K$=4 4$^+$ state, which is traditionally called  ``double $\gamma$-phonon'' state as mentioned in the previous section.  While this state is the $K$=4 rotation in the present scheme \cite{epja}, we still adopt the traditional notation $\gamma\gamma$ as 4$^+_{\gamma\gamma}$ for brevity. We derive, in the above formulation, $B(E2;4^+_{\gamma\gamma}$$\rightarrow$2$^+_{\gamma}$) / $B(E2;2^+_{\gamma}$$\rightarrow$ 0$^+_1$)=25/18$\sim$1.4, which is in agreement with known experimental trend as discussed later.
In the traditional harmonic $\gamma$-phonon model~\cite{bohr_mottelson_book2}, the intrinsic two-phonon-to-one-phonon transition strength is twice the one-phonon-to-ground-state strength. Including the rotational angular-momentum factors gives 25/9$\sim$2.8 for the same $B(E2)$ ratio. 

It is expected that the Alaga rule \cite{alaga_1955} can be derived from Eq.~(\ref{eq:Qme3}), a point to be discussed elsewhere.  The Alaga rule assumes a good $K$ quantum number, which is practically fulfilled for triaxially strongly deformed nuclei, as stated above. 

%%%%%%%%%%%%%%%%%%%%%%%%%%%%%%%

%%%%%%%%%%%%%%%%%%%%%%%%%%%%%%%

\section{Phase-specified T-plot and its application to $^{166}$Er \label{PT_166Er}}

Returning to vibrational excitations, the three bands shown in Fig.~\ref{fig:level_166Er}(b) above $1$\,MeV are the focus of present section. To analyze their nature, the characteristics of a given state, $|\Psi_k\rangle$, as defined in Eq.~(\ref{eq:mcsm_psi}), is studied based on an extension of the so-called T-plots.   

The ``Phase-specified T-plot (PT-plot)'' is presently introduced to gain information about the phase of amplitudes of basis vectors, whereas their magnitudes are considered in the usual T-plot analysis \cite{tsunoda2014,otsuka2016}.  In order to eliminate the phase ambiguity of basis states, we first choose a reference state $|\Psi_{k=R} \rangle$, in which the phase of each basis vector is fixed so that the corresponding amplitude is real positive. 
The amplitudes defining the other eigenstates are then analyzed using this set of reference basis vectors. 
Details of the actual foormulation are provided in \ref{ApB}. 
While the resulting amplitudes are in general complex numbers, they appear to be predominantly real positive or negative, for specific set of states as exemplified later.  The color filling the PT-plot circles indicates the sign of the real part of the amplitude, under the condition that the real part is sufficiently finite (its maginitude $\gtrsim$ 0.1 in this work) and the imaginary part is minor compared to it: the color is red (yellow) for a positive (negative) real amplitude. Otherwise, the color is set to white in this work, as in a single example discussed in \ref{ApB}.
%to stipulate that the real part is not significantly larger than the imaginary part in that case. 

%%%%%%%%%%%%%%%%%%%%%%%%%%%%%%%

% Fig2: 166Er  PTplot
\begin{figure}[bt]
  \centering
  \includegraphics[width=9cm]{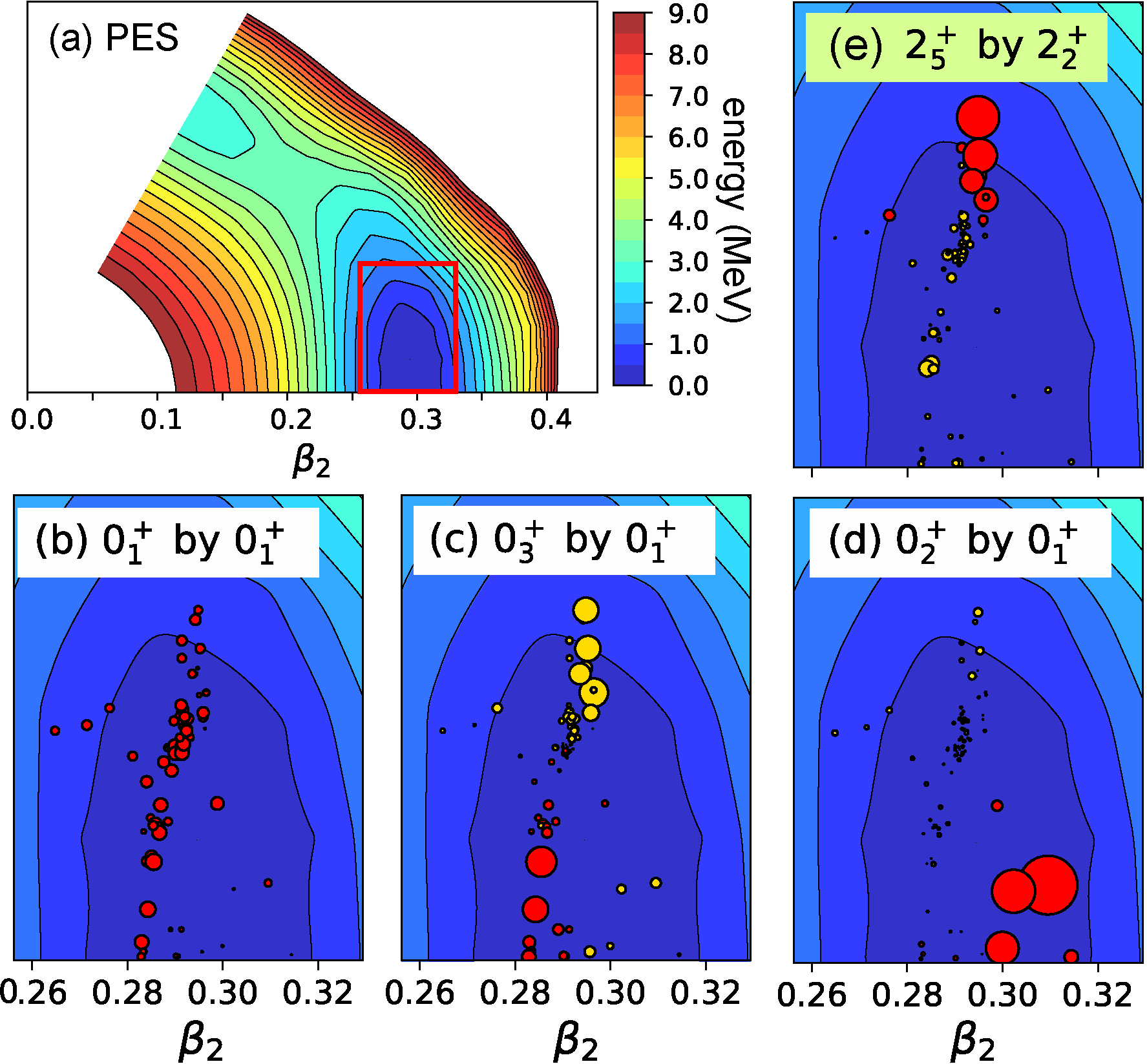}    %fig_er166_3
  \caption{(a) Particle number projected potential energy surface (PES) and (b)--(e) phase-specified T-plots for $^{166}$Er. The PT-plots focus on the part of the PES limited by the rectangle visible in panel (a). As an example of PT-plot, panel (c) depicts the 0$^+_3$ state against the 0$^+_1$ ground state chosen as a reference state (see the text). 
  }   
  \label{fig:PT_166Er}  
\end{figure}  

%%%%%%%%%%%%%%%%%%%%%%%%%%%%%%%

Figure~\ref{fig:PT_166Er}(a) displays the particle-number-projected Hartree-Fock-Bogoliubov (PNP-HFB) potential energy surface (PES) against $\beta_2$ and $\gamma$. Other panels show PT-plots within the rectangular area~\footnote{Essentially all significant contributions (i.e. circles) are located within this rectangle for all the eigenstates of present interest.} shown in panel (a), for five selected eigenstates. More eigenstates are similarly analyzed in \ref{ApB}.

By construction, all circles in panel (b) are red given that the $0^+_1$ ground state is analyzed taking itself as the reference state $|\Psi_{R} \rangle$. As visible from Fig.~\ref{fig:PT_166Er}(c), the PT-plot of the $0^+_3$ state displays a rather similar distribution of circles aligned perpendicularly to the $\gamma=0^{\circ}$ axis. However, the sign of the amplitudes alters in a characteristic way for the $0^+_3$ state. 

%%%%%%%%%%%%%%%%%%%%%%%%%%%%%%%

\begin{figure}[bt]
  \centering
%%  \vspace{0.5 cm}
  \hspace{1cm}\includegraphics[width=3cm]{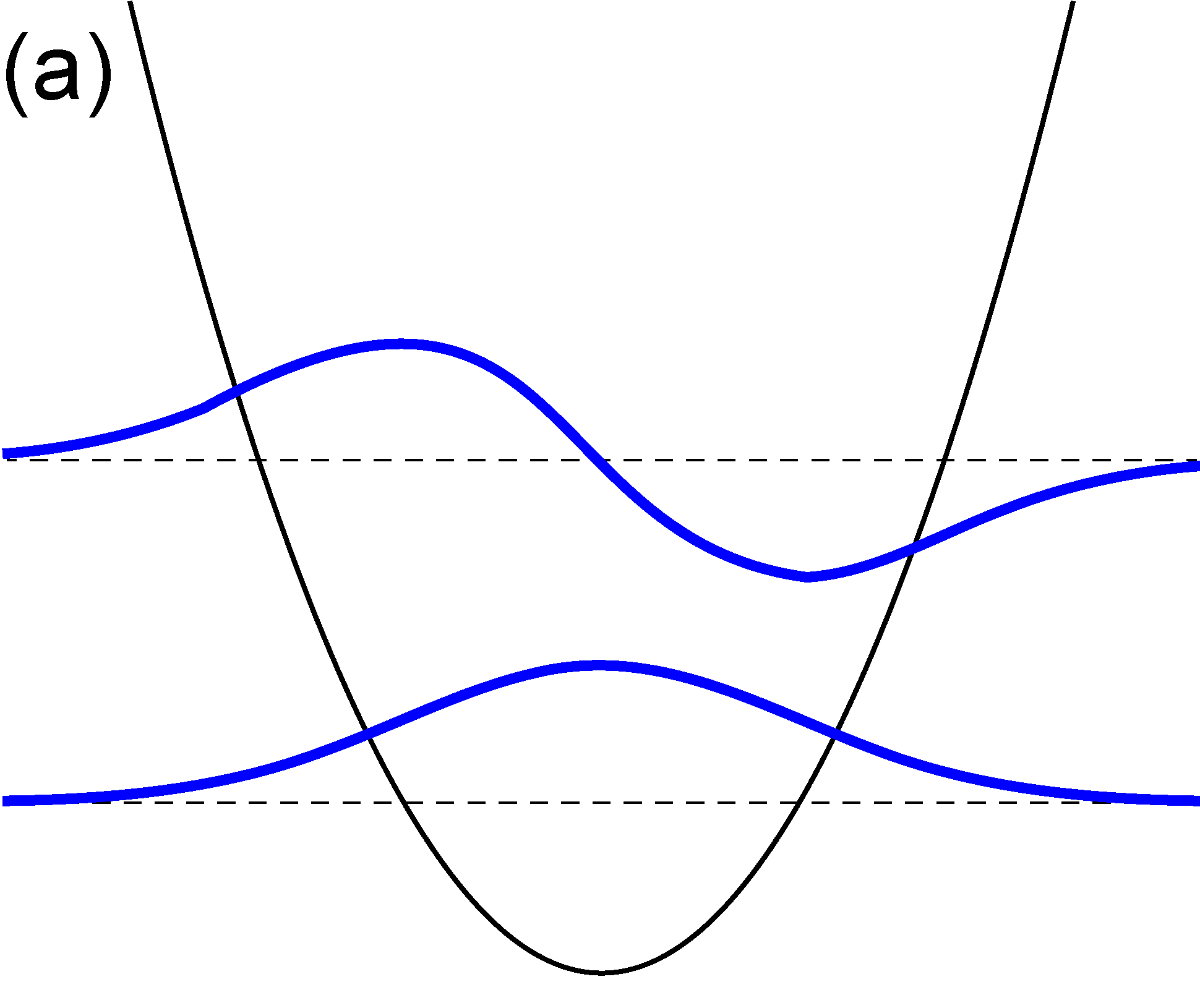}   %1dim_Ho_3
  \centering
  \includegraphics[width=8.7cm]{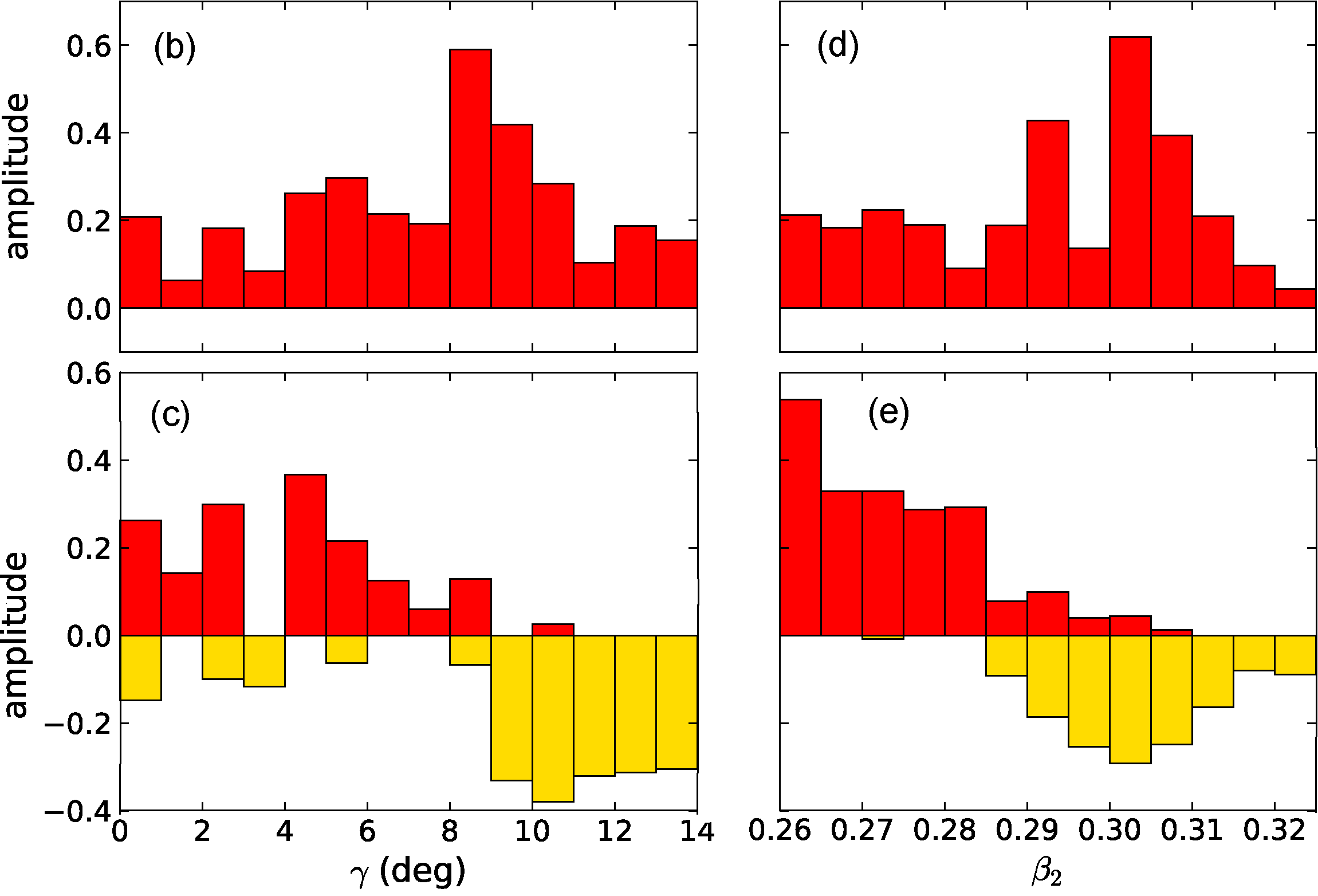}
  \caption{(a) Schematic representation of the wave functions (blue solid line) and energies (black dashed lines) of the first two eigenstates of a particle in a one-dimensional harmonic oscillator potential. 
(b)--(e) Amplitude distribution per $\gamma$ or $\beta_2$ bin. Red (yellow) histograms represent positive (negative) amplitudes for (b) $0^+_1$ and (c) $0^+_3$ states in  $^{166}$Er and for (d) $0^+_1$ and (e) $0^+_2$ states in $^{162}$Dy. }   
  \label{fig:prob_bin_sign}  
\end{figure}

To better understand such a feature, Figs.~\ref{fig:prob_bin_sign} (b) and (c) depict the amplitudes of the basis vectors per $\gamma$ bin of $1^{\circ}$ for the $0^+_1$ and 0$^+_3$ states, respectively. The linear combination of state vectors (i) belonging to a given $\gamma$ bin and (ii) having either positive or negative real amplitudes are separately superposed.  
The two vectors show positive and negative amplitudes representing the bin, as plotted by red and yellow histograms, respectively, in Fig.~\ref{fig:prob_bin_sign}. Only positive amplitudes arise for the $0^+_1$ state in panel (b), as designed.  

It is stressed that the amplitude distribution here substantially spreads with the summed probability in $\gamma$=7-9$^{\circ}$ being $\sim$56\%, which appears to be far from the basic assumptions of gamma-rigid or gamma-soft models in general.  The probability in $\gamma$=0-1$^{\circ}$ turned out to be just 4\%, driving the axial symmetry hypothesis away.  Thus, the present description may be characterized by the concept of ``shape mixing (with a strong centroid)'' like the configuration mixing concept.

Figure~\ref{fig:prob_bin_sign}(c) dedicated to the $0^+_3$ state shows that each bin contains positive and negative histograms. Very interestingly, red histograms strongly dominate the left-hand side (i.e. small $\gamma$ values) of the distribution whereas yellow ones dominate its right-hand side.   

% Fig3: infinite well + histogram

%%%%%%%%%%%%%%%%%%%%%%%%%%%%%%%

The profile of the collective amplitudes of the $0^+_1$ and $0^+_3$ states as a function of $\gamma$ recalls the behavior of the wave function of the first and second eigenstates of a particle in a harmonic oscillator potential displayed in Fig.~\ref{fig:prob_bin_sign}(a). While the ground state is a gaussian %wave packet 
function sitting in the center of the harmonic well, the wave function of the first excited state shows a change of sign at the center of the well. The comparison of panels (b) and (c) with panel (a) suggests  that  the $0^+_3$ state is itself a vibration on top of the $0^+_1$ ground state, the deformation parameter $\gamma$ playing the role of the coordinate of the potential in panel (a).  The amplitude is indeed peaked in the bin $\gamma$ = 8--9$^{\circ}$ in panel (b), whereas it is quite small in the same bin in panel (c) and displays a convincing oscillation around it. It happens that vibrational excitations with seemingly similar nodal structure can appear using, e.g., the Bohr Hamiltonian 
with triaxiality \cite{eisenberg_greiner_book1,rowe_2009,rowe_book_2010}, particularly if the potential can be approximated by a harmonic oscillator potential or a resembling one \cite{tsunoda_2021}.

 In the present work, the 0$^+_3$ state decays to the 2$^+_2$ state with relatively large $B(E2)$=9.8 W.u.
 %This is compatible with the calculated $B(E2;4^+_{\gamma\gamma} \rightarrow 2^+_2)$ = 10.9 W.u.  
This decay would be interpreted, in a traditional picture, as the decay of a double-$\gamma$-phonon state of $K^P$ = 0$^+$ \cite{fahlander_1996,garrett_1997}.  
Its calculated $B(E2)$ value, 9.8 W.u., is not far from the lower edge of an experimental value, 21$\pm$7 W.u. \cite{ensdf,garrett_1997}. Although the error is large, they share the feature of a large value.  The traditional harmonic phonon picture gives a larger value $\sim$26 W.u., because this is given by 5 $B(E2;2^+_{\gamma} \rightarrow 0^+_1)$. The E2 decay to the 2$^+_1$ state seems to be undetectably small \cite{garrett_1997} consistently with the calculated $B(E2)$=0.3 W.u. (consistent with the Alaga rule) (see Fig.~\ref{fig:level_166Er}).

The E2 transition associated with a $\gamma$ vibrational mode can be somewhat stronger than single-particle estimates thanks to the fact that the E2 transition between triaxial states increases with $\gamma$ according to the discussion of Sec.~\ref{sec:triaxial} such that significant amplitudes at large $\gamma$ contributing to the vibrational state help building up relatively strong collective transitions. The same trend is true for $\beta_2$ vibrations, as we shall see next.

One may wonder why the $\gamma$ vibration displays a $0^+$ band head, rather than a $2^+$ band head {\it \`a la} A. Bohr.  The present $\gamma$ vibrational states belong to a set of triaxial states 
characterized by the pattern like Fig.~\ref{fig:prob_bin_sign}.  
The states differing only in $K$ quantum numbers are split in energy by the mechanism explained based on Fig. 6 of \cite{epja}: the $K^P$ = 0$^+$ band is robustly lower than the $K^P$ = 2$^+$ band.

The PT-plot of the $0^+_2$ state with $E_x\sim$ 2 MeV in Fig.~\ref{fig:PT_166Er}(d) suggests that it is more deformed ($\beta_2$$\sim$0.31) than the $0^+_{1,3}$ states ($\beta_2$$\sim$0.29) and a good candidate for a shape-coexistence case with respect to $\beta_2$ \cite{wood1992,andreyev2000,heyde2011,leoni2024}.  
Figure~\ref{fig:level_166Er} includes $B(E2;0^+_2\rightarrow2^+_{1,2})$. The decay to the 2$^+_2$ shows good agreement between theory and experiment.  The experimental values are calculated from the lifetime and branching ratios depicted in \cite{nudat3}, and other possible  contributions are not included.  Despite this problem, this data appears to be supportive of the present discussion.  The decay to the 2$^+_1$ shows that the theoretical value is  a few factor smaller than the experimental value \footnote{The 2$^+_1$ state is connected by a large E2 matrix element to the 0$^+_1$ state, and a tiny 0$^+_1$-0$^+_2$ mixing can change small $B(E2;0^+_2 \rightarrow 2^+_1)$ value.  This is a common feature for the decay of excited $0^+$ states to the 2$^+_1$ state.}.

In order to analyze the $2^+_2$ state, we cannot use the $2^+_1$ state as the reference state, because the former is the lowest $K^P$ = 2$^+$ rotational excitation of the triaxial ground state as discussed in Ref.~\cite{epja}, whereas the latter is a member of the lowest $K^P$ = 0$^+$ band. It is thus more appropriate to analyze the $2^+_2$ state against itself as shown in \ref{ApB}.  
\textcolor{black}{
With this at hand, Fig.~\ref{fig:PT_166Er}(e) shows that, just like the $0^+_3$ state is the $K^P$ = 0$^+$ $\gamma$ vibration on top of the $0^+_1$ state, the 2$^+_5$ state results from a $\gamma$ vibrational excitation on top of $2^+_2$ state.  Note that the $0^+_1$ and $2^+_2$ states belong to the same family, just differing in their $K$ value.}
%is analyzed by the reference state $2^+_2$, exhibiting as  a $\gamma$ vibrational excitation .   

\textcolor{black}{Eventually,} real amplitudes distinctly emerge as a signature of vibrational excitations, while amplitudes can be complex numbers in other states. %as demonstrated in \ref{ApB}.
Remarkably, no vibrational states in the $\beta_2$ direction have been found for $^{166}$Er in the excitation-energy range under scrutiny.

\textcolor{black}{The $4^+$ state labeled as ``$\gamma\gamma$ band ($K^P$=4$^+$)'' in Fig.~\ref{fig:level_166Er} (b) depicts a PT-plot similar to the one for the $2^+_2$ state (see \ref{ApB}), which is consistent with its $K^P$=4$^+$ rotational picture and further with \cite{epja}.   Experimentally $B(E2;4^+_{\gamma\gamma}\rightarrow$ 2$^+_2$) is 8$\pm$3 W.u. in the ensdf \cite{ensdf}, in agreement with the calculated one, 10.9 W.u. (see Fig.~\ref{fig:level_166Er}). 
% whereas the traditional harmonic phonon model leads to 14.4$\pm$0.6 W.u. due to the ratio, 2.78 (see Sec.~\ref{sec:triaxial}).  
} 

\textcolor{black}{
The MCSM values give $B(E2;4^+_{\gamma\gamma}$$\rightarrow$2$^+_2$) / $B(E2;2^+_2\rightarrow$ 0$^+_1$)= 2.1, compared to 1.4 in the simple modeling discussed in Sec.~\ref{sec:triaxial}. This is probably because $\gamma$ becomes gradually larger with $K$, as a stretching effect \cite{epja}. This ratio obtained from the ensdf values is in the range of 1.0-2.2. The present value is within this range, whereas the harmonic phonon yields 2.78 (see Sec.~\ref{sec:triaxial}).
} 

\textcolor{black}{
These two transitions appear to be more consistent with the present triaxiality view than with the harmonic phonon model, considering the energies also.  Further studies are of interest.
}

%%%%%%%%%%%%%%%%%%%%%%%%%%%%%%%%%%%%%%%%%%%%%%

%%%%%%%%%%%%%%%%%%%%%%%%%%%%%%%%%%%%%%%%%%%%%

\section{Application to $^{162}$Dy with $\beta$-vibrational band emergence}

%%%%%%%%%%%%%%%%%%%%%%%%%%%%%%%

% Fig4: 162Dy
\begin{figure}[bt]
  %\centering
  \includegraphics[width=9cm]{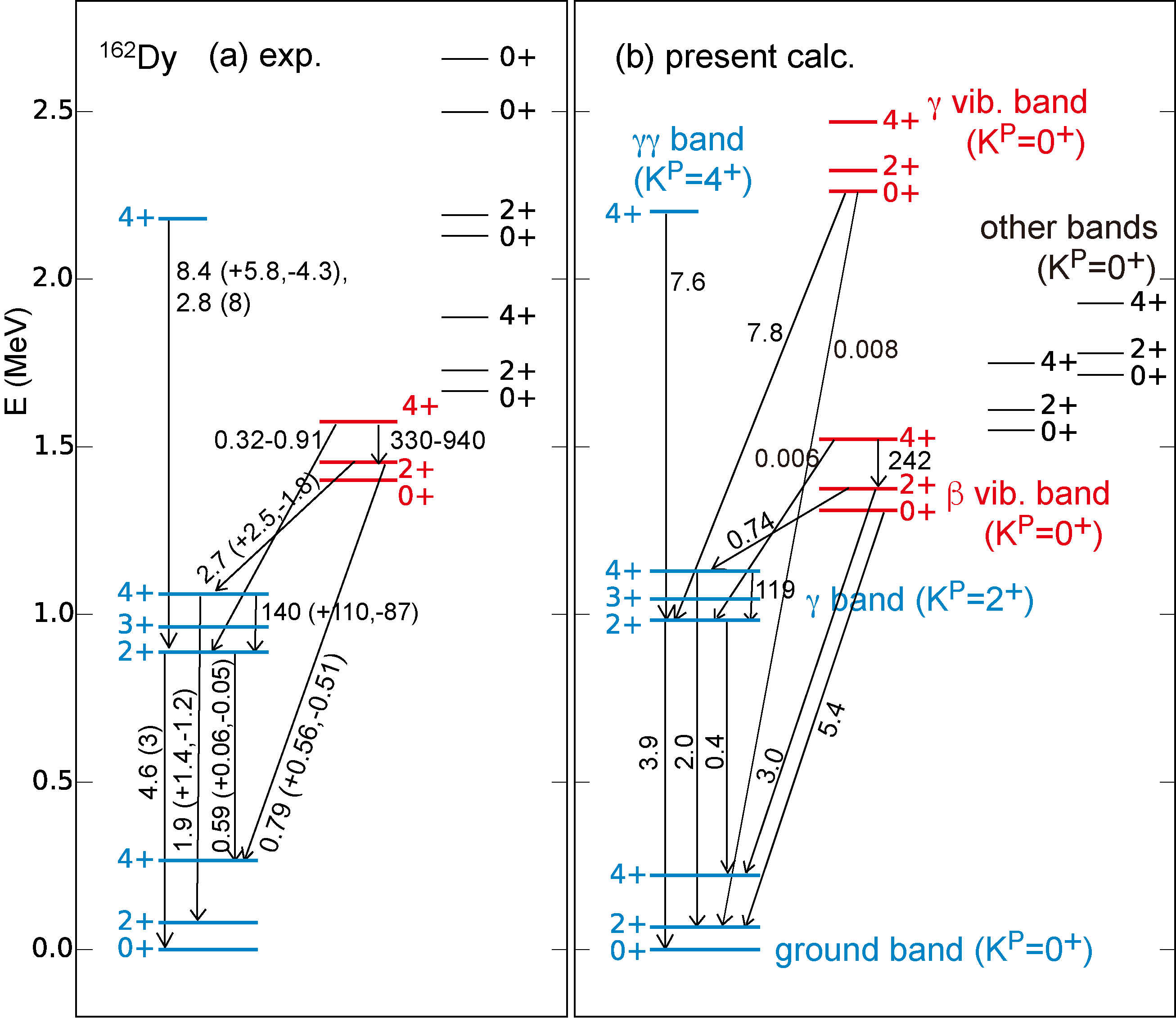}  %162Dy_levels_4red
  \caption{Same as Fig.~\ref{fig:level_166Er} for $^{162}$Dy.  Experimental data are from \cite{ensdf,aprahamian_2024}. 
  }   
  \label{fig:level_162Dy}  
\end{figure}  
%%%%%%%%%%%%%%%%%%%%%%%%%%%%%%%%%

Moving now to $^{162}$Dy, Fig.~\ref{fig:level_162Dy} depicts experimental and calculated energies of low-lying positive-parity states with angular momenta $J=0, 2, 3, 4$, along with selected $B(E2)$ values. 
Experimental $B(E2)$ values are taken from Refs.~\cite{ensdf,aprahamian_2024}, avoiding cases with E2/M1 mixing.  If just boundaries, lower or upper, are given, they are not shown.
The overall agreement between the present calculation and experiment is remarkable.

%%%%%%%%%%%%%%%%%%%%%%%%%%%%%%%

% Fig5: 162Dy  PTplot
\begin{figure}[bt]
  \centering
    \includegraphics[width=8.9cm]{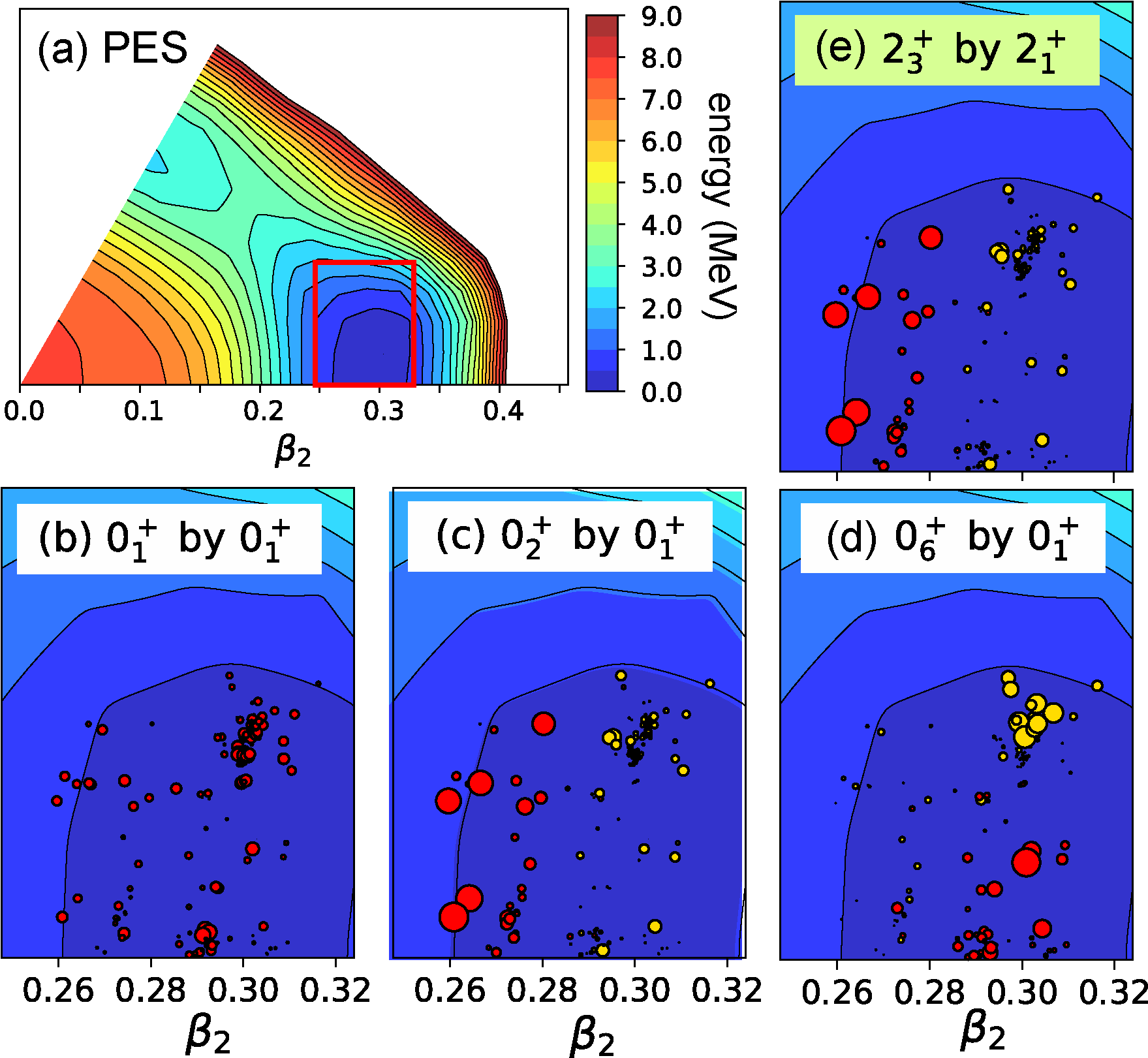}  %fig_dy162_2

  \caption{
  Same as Fig.~\ref{fig:PT_166Er} for selected states in $^{162}$Dy.
  }   
  \label{fig:PT_162Dy}  
\end{figure}  

%%%%%%%%%%%%%%%%%%%%%%%%%%%%%%%

Similarly to $^{166}$Er, Fig.~\ref{fig:PT_162Dy} displays the PNP-HFB PES and the PT-plots for a selected set of states.
Figure~\ref{fig:PT_162Dy}(c) depicts the PT-plot of the $0^+_2$ state composed of red and yellow symbols.  
\textcolor{black}{This is nothing but a vibrational excitation in the $\beta_2$ direction, which is well verified by the nice agreement between theory and experiment for energy levels and E2 transitions.   
Figure~\ref{fig:prob_bin_sign} (d) and (e) demonstrate the vibrational structure with an analysis similar to   Fig.~\ref{fig:prob_bin_sign} (b), (c), with the coordinate being $\beta_2$ instead of $\gamma$.} Again, the $0^+_1$ ground state and the $0^+_2$ excited state conform well to the expected picture of a harmonic oscillator, the latter behaving as a low-lying $\beta$-vibrational excitation on top of the former~\footnote{The distribution of weights being asymmetric around the node, the vibration is somewhat anharmonic. This is consistent with the behavior of the PES in the $\beta_2$ direction.}. This is made possible by the wider bottom of the PES in the $\beta_2$ direction for $^{162}$Dy than for $^{166}$Er. 
%It is of interest to see if future experiments confirm the appearance of such a $0^+$ state.

Figure~\ref{fig:PT_162Dy}(d) indicates that the $0^+_6$ state carries a $\gamma$ vibration on top of the $0^+_1$ state. 

As seen from Fig.~\ref{fig:PT_162Dy}(e), the $2^+_3$ state presents the same $\beta$-vibrational structure as the $0^+_2$ state (see Fig.~\ref{fig:PT_162Dy}(c)), i.e. it belongs to the rotational band built on top of the latter.

The $4^+_7$ state is the $J^P$ = $K^P$ = 4$^+$ rotation of
the almost same triaxial intrinsic state as the ground and $\gamma$ bands (see \ref{ApB}).  
A recent experiment \cite{aprahamian_2024} indicates that the $B(E2;4^+_{\gamma\gamma} \rightarrow 2^+_2)\sim$ 8.4 W.u. is in a good agreement with the present calculation (see Fig.~\ref{fig:level_162Dy}(b)).

%%%%%%%%%%%%%%%%%%%%%%%%%%%%%%%%%%%%%%%%%%%%%%

%%%%%%%%%%%%%%%%%%%%%%%%%%%%%%%%%%%%%%%%%%%%%

\section{Summary}

A novel picture of vibrational modes in strongly deformed heavy nuclei, at excitation energies below a few MeV, was presented in the present work. 
\textcolor{black}{The concept of the prevailing triaxiality in heavy deformed nuclei is reinforced by deriving E2 transition properties for triaxial states.  The obtained ${\rm sin}^2 \gamma$ dependence of the $B(E2;0^+_1 \rightarrow 2^+_2)$ transition reproduces the trend in heavy strongly-deformed nuclei.}  Besides triaxial states forming the ground, $\gamma$, and $\gamma\gamma$ bands, a variety of vibrational bands generally appear.  Their excitation energies are higher in most cases, but the $\beta$ band of $^{162}$Dy appears rather low, in good agreement with experiment.  In the same energy region as such vibrational excitations, shape coexistence bands with similar $\beta_2$ values also appear, many of which may be of low excitation energies largely due to the self-organization  \cite{otsuka_2019}.   The $0^+$ state belonging to the $\gamma$-vibrational band decays rather strongly to the $2^+_{\gamma}$ state.  This can be confused as the decay of a double $\gamma$ phonon $0^+$ state, but it is a single vibrational phonon excitation.  Due to triaxiality, a $K^P$ = 0$^+$ vibrational band appears at lower energy than the corresponding $K^P$ = 2$^+$ vibrational band.  Thus, a new comprehensive vision over quadrupole collective states may be emerging, and eventually, a number of intriguing phenomena, such as more $K$ rotations, vibrational bands, shape coexistence with $\beta_2$ or $\gamma$, will provide fruitful agendas for carrying new nuclear spectroscopy studies.   It may be worth noting that despite substantial differences in the structures of the two example nuclei, both of them are nicely described by the present QVSM calculation with the same Hamiltonian. 

%%%%%%%%%%%%%%%%%%%%%%%%%%%%%%%%%%%%%%%%
The authors are grateful to Dr. P. Van Duppen for valuable suggestions at early stage of the work, to Dr. A. G\"orgen and Dr. M. A. Caprio for useful comments on vibrational modes, and to Dr. H. Ueno for continuous support to this research activity.    TO thanks the visitor program of GANIL and the Alexander von Humboldt Foundation for the Research Award, as some parts of this work were made under their supports.  TO is grateful to Dr. T. Kobori for encouraging remarks.  The QVSM calculations were performed on the supercomputer Fugaku at RIKEN R-CCS  (hp190160, hp200130, hp210165, hp220174, hp230207, hp240213, hp250224).  
This work was supported in part by MEXT as ``Program for Promoting Researches on the Supercomputer Fugaku'' (Simulation for basic science: from fundamental laws of particles to creation of nuclei, JPMXP1020200105, Simulation for basic science: approaching the new quantum era, JPMXP1020230411), and by JICFuS.  
This work was supported by JSPS KAKENHI Grant Numbers JP24K07043 and JP25K00998.  
%This work was supported by JSPS KAKENHI Grant Numbers JP21K03564, JP20K03981 and JP18H05462.  

%%%%%%%%%%%%%%%%%%%%%%%%%%%%%%%%%%%%%%%%%%%%%%
%\makeatletter
%%%%%%%%%\renewcommand\@biblabel[1]{#1.}
%\renewcommand\@cite[2]{\ts{#1\if@tempswa , #2\fi}}
%%%%%%%%%%%\makeatother
%%%%%%%%%%\newcommand{\bibitemnature}[1]{\addtocounter{enumii}{1}\bibitem[\number\value{enumii}]{#1}} %% To use numbers for methods summary also
%%%%%%%%%%\def\bibsection{\section{R\lowercase{eferences}}}
%%%%%%%%%%

%%%%%%%%%%%%%%%%%%%%%%%%%%%%%%%
%%%%  A P P E N D I X   %%%%

\appendix

\section{E2 transition between triaxial states \label{ApA}} 

In Appendix A, we present some details of the mathematical manipulations in Sec.~\ref{sec:triaxial}.  For the $D$ function and other related quantities, we adopt the convention of Edmonds \cite{edmonds}, which is also used in \cite{ring_schuck_book} with a summary in its Appendix A.  

The angular-momentum projected state obtained from $ | \, \phi \rangle$ is written as,
\begin{eqnarray}
\lefteqn{ | J, M, K \rangle }\nonumber \\
&\,=&\frac{2J+1}{8 \pi^2} \int_0^{2\pi}   d\alpha \int_0^{\pi} d\beta \, {\rm sin}\beta \, \int_0^{2\pi} d\gamma \nonumber \\
&& \big\{ D^J_{MK}(\alpha, \beta, \gamma) \big\}^*\, {\hat R}(\alpha, \beta, \gamma) \, |\phi \rangle , \nonumber \\
&\,=&\frac{2J+1}{8 \pi^2} \int   d\bm{\omega} \,[D^J_{MK} (\bm{\omega})]^* \, {\hat R}(\bm{\omega}) \, | \phi \rangle ,
\label{eq:rot_phi}
\end{eqnarray}
where $\bm{\omega}$ collectively implies Euler angle $(\alpha, \beta, \gamma)$, the rotation operator is defined as,
\begin{equation}
{\hat R}(\bm{\omega}) \,=\, {\hat R}(\alpha, \beta, \gamma) \,=\, e^{i\alpha \hat{J}_z} \, e^{i\beta \hat{J}_y} \, e^{i\gamma \hat{J}_z} ,
\label{eq:R}
\end{equation}
$D$ stands for the Wigner's $D$ function, 
\begin{equation}
D^J_{MK} (\bm{\omega})\,=\,D^J_{MK} (\alpha, \beta, \gamma) \,=\, e^{iM\alpha} \, d^J_{MK} (\beta) \, e^{iK\gamma } ,
\label{eq:Wigner}
\end{equation}
with $d^J_{MK} (\beta)$ being the (small) $d$ function.  

The following operator in eq.~(\ref{eq:Qme}), 
\begin{eqnarray}
&\lefteqn{{\hat R}^{-1}(\bm{\omega}) \, Q_{2,\mu}  \, {\hat R}(\bm{\omega})} \nonumber \\
&\,\,\,\, = &e^{-i\gamma \hat{J}_z} \, e^{-i\beta \hat{J}_y} \, e^{-i\alpha \hat{J}_z} \, Q_{2,\mu}  \, e^{i\alpha \hat{J}_z} \, e^{i\beta \hat{J}_y} \, e^{i\gamma \hat{J}_z},
\label{eq:Q_der1}
\end{eqnarray}
can be rewritten, because $e^{-i\alpha \hat{J}_z}\, Q_{2,\mu}  \, e^{i\alpha \hat{J}_z}$ can be expressed as a c-number phase factor $e^{-i\alpha \mu}$ ($J_z$ value is changed by $\mu$) in front of $Q_{2,\mu}$, and then because $e^{-i\beta \hat{J}_y} \, Q_{2,\mu}  \, e^{i\beta \hat{J}_y} $ is given by the $d^2_{\nu\mu}(-\beta)$ functions.   
The consequence is shown below, and is further processed as,
\begin{eqnarray}
&\lefteqn{e^{-i\alpha \mu} \, e^{-i\gamma \hat{J}_z} \, \{\sum_{\nu} \, d_{\nu\mu}(-\beta)\,Q_{2,\nu} \} \, e^{i\gamma \hat{J}_z}}  \nonumber \\
&\,\,\,\, = &e^{-i\alpha \mu} \, \{\sum_{\nu} \, e^{-i\gamma \nu} \, d_{\mu\nu}(\beta)\,Q_{2,\nu}  \}\nonumber \\
&\,\,\,\, = &\sum_{\nu} \, e^{-i\alpha \mu} \, d_{\mu\nu}(\beta) \,e^{-i\gamma \nu} \,Q_{2,\nu}  \nonumber \\
&\,\,\,\, = &\sum_{\nu} \, [D^2_{\mu\nu} (\alpha, \beta,\gamma)]^*  \,Q_{2,\nu}  ,
\label{eq:Q_der}
\end{eqnarray}
where the phase factor, $e^{-i\gamma \nu}$, appears similarly to $e^{-i\alpha \mu}$ in eq.~(\ref{eq:Q_der1}) and 
\begin{equation}
d^J_{\nu \mu} (-\beta) \,=\, d^J_{\mu \nu} (\beta),
\label{eq:d_numu}
\end{equation}
is used.

The last part of eq.~(\ref{eq:Qme5}) is derived with the relations,
\begin{equation}  
\int d\bm{\omega} \,[D^J_{MK} (\bm{\omega})]^* \, D^{J'}_{M'K'} (\bm{\omega})  %} \nonumber \\
= \frac{8\pi^2}{2J+1} \delta_{JJ'} \delta_{MM'} \delta_{KK'} , %}
\label{eq:DDint}
\end{equation}  
and
\begin{eqnarray}
\lefteqn{D^k_{\mu\nu} (\bm{\omega}) \, D^{J}_{MK} (\bm{\omega}) } \nonumber \\
&=& \sum_{J'}  (J M k \mu | J' \,M+\mu) \, (J K k \nu | J' \,K+\nu) \, D^{J'}_{M+\mu \,K+\nu}(\bm{\omega}).\nonumber \\ 
\label{eq:DDproduct}
\end{eqnarray}

%%%%%%%%%%%%%%%%%%%%%%%%%%%%%%%%%%%%%%%%%%%%%
\section{PT-plot \label{ApB}}

In Appendix B, we provide detailed discussions about the PT-plot as well as more results of the analyses by it. 

As the basis vectors $|\phi_{iK}\rangle$ in eq.~(\ref{eq:mcsm_psi}) 
are non-orthogonal among themselves, their linear combinations are taken to be orthonormal, by diagonalizing the norm matrix, resulting in the states $| \tilde{\phi}_{iK}\rangle$ comprising linear combinations of $|\phi_{i'K'} \rangle$ with $i' = 1, 2, \ldots$ and $K' = -J, ..., J$.  
The eigenstate labeled by $k$ is written as,
\begin{equation}
 |\Psi_k\rangle=\sum_{i} \, \Biggr\{ \,\sum_{K} \, \tilde{f}^{(k)}_{iK} |\tilde{\phi}_{iK}\rangle \, \Biggl\} \, , 
\label{eq:mcsm_psi_orth}
\end{equation}
where $\tilde{f}^{(k)}_{iK}$ denotes the orthonormalized amplitudes obtained by diagonalizing the Hamiltonian in the Hilbert space spanned by $| \tilde{\phi}_{iK}\rangle$'s. The amplitudes $\tilde{f}^{(k)}_{iK}$'s are taken to be all real positive for $k=R$ ($R$: reference state) (see Sec.~\ref{sec:triaxial}).

We introduce $K$-integrated normalized states for the index $i$,
\begin{equation}
| \tilde{\phi}^{(k)}_{i}\rangle =\frac{1}{{\cal N}^{(k)}_i} \,  \sum_{K}  \tilde{f}^{(k)}_{iK}  \, | \tilde{\phi}_{iK}\rangle \, , 
\label{eq:psiK}
\end{equation}
where ${\cal N}^{(k)}_i$ is a normalization constant set to be real positive and the relation, 
\begin{equation}
\langle \tilde{\phi}^{(k)}_{i} | \tilde{\phi}^{(k)}_{i'}\rangle \,=\, \delta_{i,i'}, 
\label{eq:psiK_delta}
\end{equation}
is obvious.  In the visualization of the amplitudes, we adopt these $K$-integrated basis vectors. Mainly because the $K$ distribution for a specific $i$ in eq.~(\ref{eq:psiK}) is not widely spread and is rather concentrated to natural components at least for collective states,    
it is a sensible choice to expand a general eigenstate in terms of the states in eq.~(\ref{eq:psiK}) with $k$=$R$.  The expansion becomes
\begin{equation}
 |\Psi_k\rangle=\sum_{i} \, | \tilde{\phi}^{(R)}_{i}\rangle \, {\cal N}^{(k)}_i \langle \tilde{\phi}^{(R)}_{i} | \tilde{\phi}^{(k)}_{i}\rangle + {\rm rest } ,
\label{eq:psiK_expansion}
\end{equation}
where ``rest'' means the rest component.   This rest component is supposed to be minor, because both $| \tilde{\phi}^{(R)}_i \rangle$ and $| \tilde{\phi}^{(k)}_i\rangle$ stem from the same $| \phi_{i}\rangle$ with similar $K$ distributions. Because their amplitudes in a given eigenstate $|\Psi_k\rangle$ can be different, implying that the overall magnitudes before the normalization by ${\cal N}^{(k)}_i$ generally vary.  We point out that $K$ takes only zero for $J$=0 in the summation in eqs.~(\ref{eq:mcsm_psi_orth}) and (\ref{eq:psiK}), resulting a vanished rest component.
We later examine this truncation by a concrete numerical calculation.

We consider the amplitude given by    
\begin{equation}
X^{(k)}_i \,=\, \langle \tilde{\phi}^{(R)}_{i} | \Psi_k\rangle \,=\,  {\cal N}^{(k)}_i \langle \tilde{\phi}^{(R)}_{i} | \tilde{\phi}^{(k)}_{i}\rangle ,
\label{eq:psiK_ampl}
\end{equation}
where ${\cal N}^{(k)}_i$ expresses the probability amplitude of the $i$-th basis vector, $| \tilde{\phi}^{(k)}_{i}\rangle$, in the eigenstate $k$, and 
$\langle \tilde{\phi}^{(R)}_{i} | \tilde{\phi}^{(k)}_{i}\rangle$ represents the overlap of the two states normalized.

The remaining problem is to fix the overall phase of $| \tilde{\phi}^{(k)}_{i}\rangle$ for $k \ne R$.  This phase is determined so that the overlap, $X^{(k)}_i$, becomes a real positive number for $i$ giving the maximum of ${\cal N}^{(R)}_i {\cal N}^{(k)}_i | \langle \tilde{\phi}^{(R)}_{i} | \tilde{\phi}^{(k)}_{i}\rangle |$.
We thus fix the amplitudes $X^{(k)}_i$ up to their phases.  The amplitudes $X^{(k)}_i$ are used to draw Fig.~\ref{fig:prob_bin_sign}. 
\\

%%%%%%%%%%%%%%%%%%%%%%%%%%%%%%%%%%%%%%%%%%%%%%%%%%%

Given a physical eigenstate $|\Psi_k\rangle$, the T-plot displays at the value of the deformation parameters $\beta_2$ and $\gamma$ of each basis vector $| {\phi}^{(i)}\rangle$  
a circle whose area exhibits the $K$-summed probability $|{\cal N}^{(k)}_i|^2=\sum_{K} |\tilde{f}^{(k)}_{iK}|^2$     
\footnote{The T-plot  actually uses the $\beta_2$ and $\gamma$ values computed for the basis vector $|\phi\rangle$ \cite{utsuno_2015,otsuka_2022}. Because the mixing due to the orthogonalization links two such states with very close values of the deformation parameters, $| \tilde{\phi}_{iK}\rangle$ can refer to the deformation parameters of $| \phi_{iK}\rangle$.}.  
The T-plot was originally introduced for non-orthogonal basis vectors~\cite{tsunoda2014,otsuka2016}, but is presently made with the orthonormality \cite{otsuka_2022},  in order to incorporate finer details of wave functions.
The phase of the amplitude can be further shown now by different color for each circle of the PT-plot.  As shown in Figs.~\ref{fig:PT_166Er} and \ref{fig:PT_162Dy}, the T-plot circle dominated by positive (negative) real part is painted in red (yellow).  If the real and imaginary parts are finite and comparable, a weaker color can be used, although this is not the case in this paper.  Otherwise, the color is set to be white, including the cases that the $K$-distribution substantially differs between the two eigenstates. 

%%%%%%%%%%%%%%%%%%%%%%%%%%%%%%%

% Fig6: 166Er  PTplot   supplementary
\begin{figure}[bt]
  \centering
  \vspace{0.5cm}
  \includegraphics[width=8.9cm]{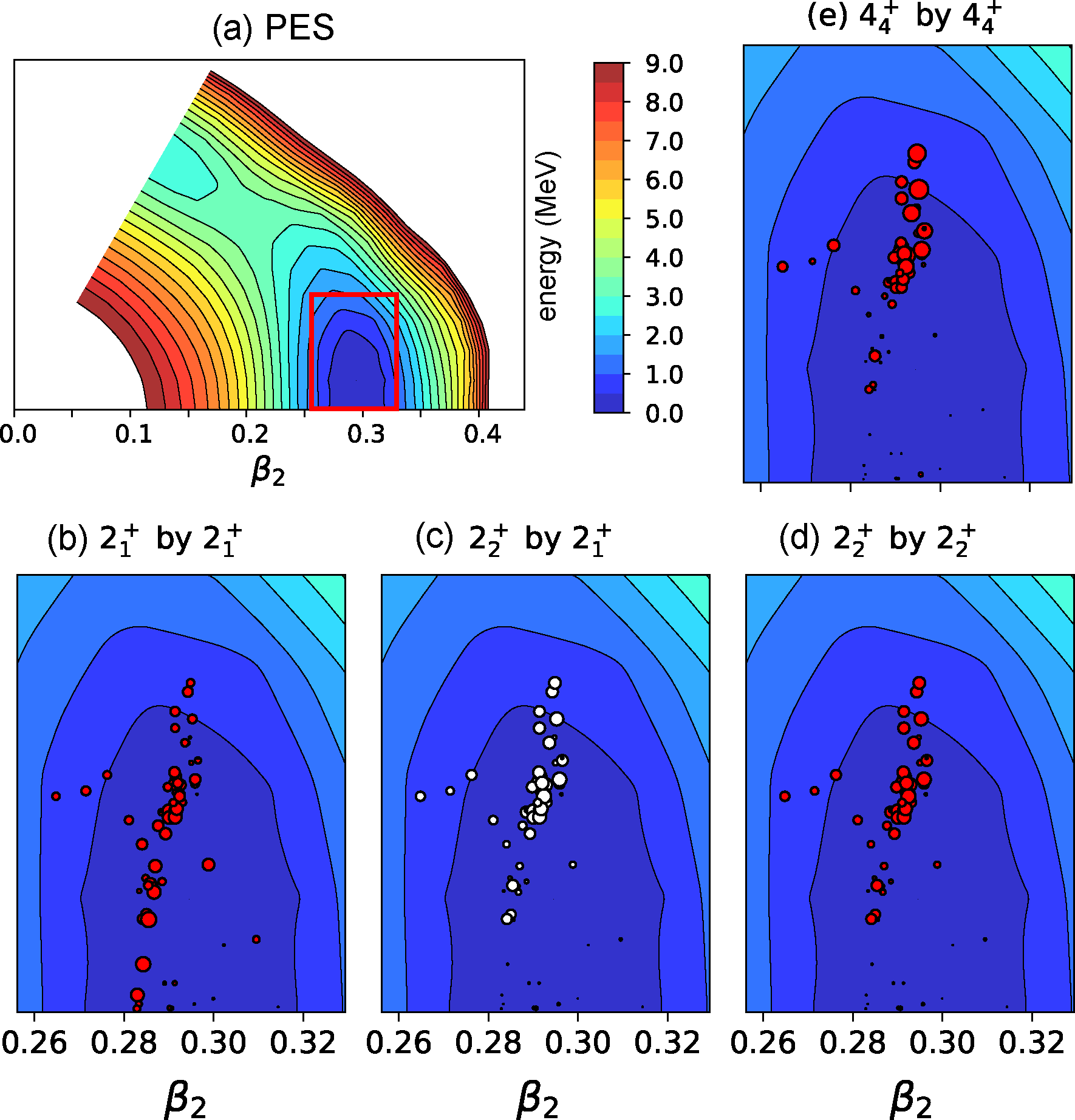}    %fig_er166_3
  \caption{Phase-specified T-plots for $^{166}$Er  for some selected states.  See the caption of Fig.~\ref{fig:PT_166Er}.    }   
  \label{fig:PT_166Er_AP}  
\end{figure}  

%%%%%%%%%%%%%%%%%%%%%%%%%%%%%%%

Figure~\ref{fig:PT_166Er_AP} displays additional PT-plot for $^{166}$Er.
The PT-plot of the $2^+_1$ state in panel (b) is very similar to that of the $0^+_1$ state in Fig.~\ref{fig:PT_166Er} (b). This is related to the fact that this state is the $2^+$ excitation inside the ground-state rotational band.
The white circles in Fig.~\ref{fig:PT_166Er_AP} (c) confirms that the 2$^+_1$ state should not be used as a reference state to describe the $2^+_2$ state, because the 2$^+_1$ (2$^+_2$) state is basically a pure $K$=0 (2) state. 
The $2^+_2$ state is then analyzed against itself as shown in Fig.~\ref{fig:PT_166Er_AP} (d).

Panel (e) displays the PT-plot of the $4^+_4$ state similar to panel (d) with a slight overall shift towards stronger triaxiality, confirming its $J^P$ = $K^P$ = 4$^+$ character \cite{epja}.  

\begin{figure}[bt]
  \centering
    \vspace{0.5cm}
  \includegraphics[width=8.9cm]{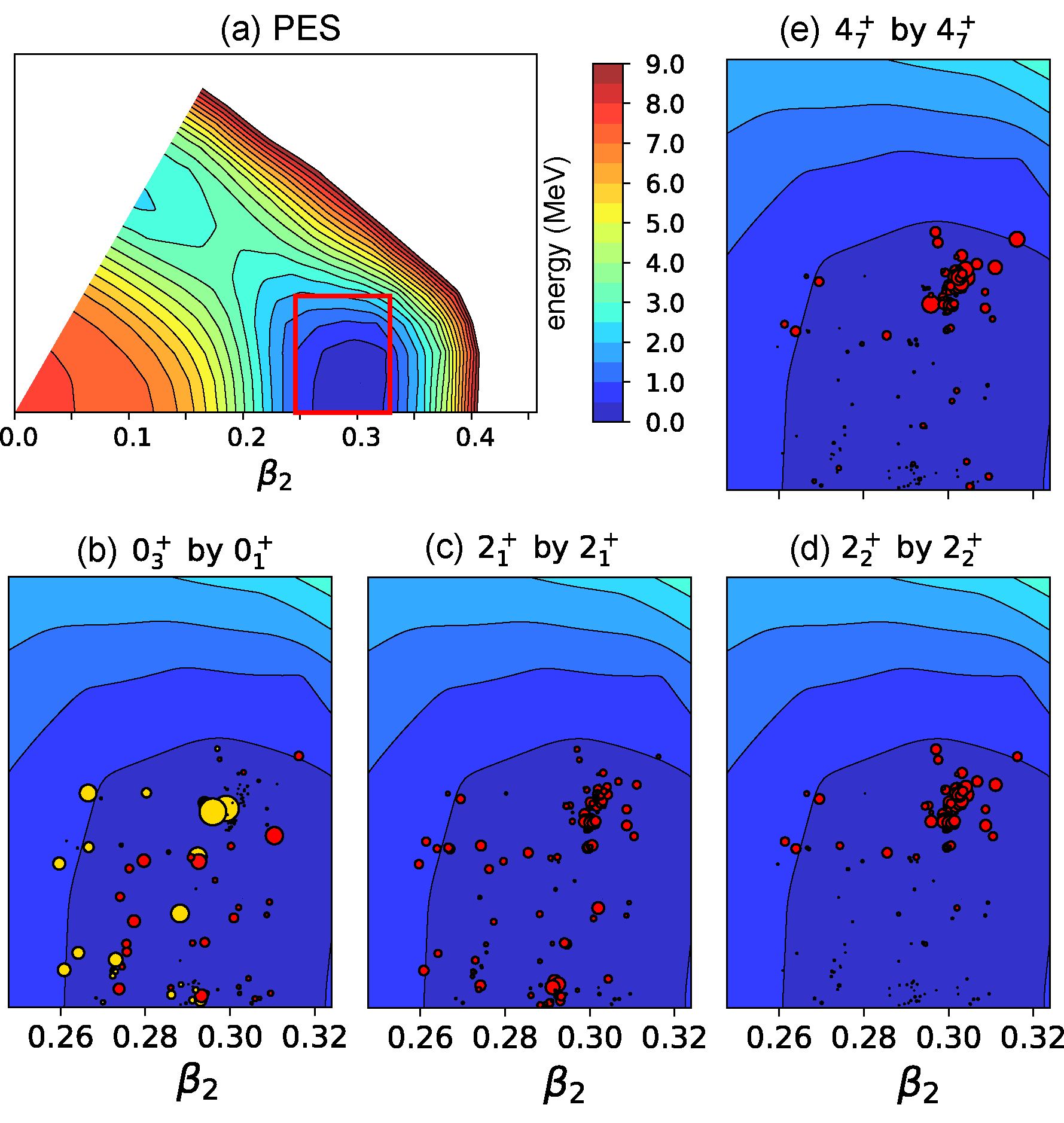}    %fig_er166_3
  \caption{Phase-specified T-plots for $^{162}$Dy  for some selected states.  See the caption of Fig.~\ref{fig:PT_162Dy}.    }   
  \label{fig:PT_162Dy_AP}  
\end{figure}

The PT-plot for $^{162}$Dy is shown in Fig.~\ref{fig:PT_162Dy_AP}.
Figure~\ref{fig:PT_162Dy_AP}(b) shows the PT-plot for the $0^+_3$ state, indicating that the superposition of various basis vectors occurs in a much more mixed way.   

%%%%%%%%%%%%%%%%%%%%%%%%%%%%%%%

% Fig6: 166Er  PTplot   supplementary

%%%%%%%%%%%%%%%%%%%%%%%%%%%%%%%

Figure~\ref{fig:PT_162Dy_AP}(c) displays the PT-plot for the $2^+_1$ state, which is almost identical to Fig.~\ref{fig:PT_162Dy}(b).   Figure~\ref{fig:PT_162Dy_AP}(d) indicates
that the triaxiality of the $2^+_2$ state is slightly enlarged compared to the $2^+_1$ state, consistently with the argument in \cite{epja}.  

Figure~\ref{fig:PT_162Dy_AP}(e) confirms by means of the PT-plot that the $4^+_7$ state is the $J^P$ = $K^P$ = 4$^+$ rotation of the almost same triaxial intrinsic state as the ground and $\gamma$ bands.  

Referring to the cases shown above, we examine the truncation made in eq.~(\ref{eq:psiK_expansion}).
Among the 16 cases shown in Figs.~\ref{fig:PT_166Er}, \ref{fig:PT_162Dy}, \ref{fig:PT_166Er_AP}, \ref{fig:PT_162Dy_AP}, eight are for the combinations where the reference state differs from the state of interest: if they are equal the truncation is exact. Figure ~\ref{fig:PT_166Er}(e) exhibits the case for the 2$^+_5$ state against the 2$^+_2$ state ($K$=2) in $^{166}$Er. The former is the band head of $\gamma$ vibrational $K$=2 band, and the truncation fills the probability up to 90.0\% with real parts only.  An increase of 0.6\% is gained by including imaginary parts.  Considering the higher excitation energy ($\sim$4 MeV), this seems to indicate an amazing collectivity with a vibrational character.  A nearly perfect truncation is found in Fig.~\ref{fig:PT_162Dy}(e), where the 2$^+_3$ state ($K$=0 or a $\beta$-band member) in $^{162}$Dy is analyzed with the 2$^+_1$ state ($K$=0).  The truncation reaches 98.5\% with real parts, and 99.8\% including imaginary parts.
In contrast to these cases, Fig.~\ref{fig:PT_166Er_AP}(c) exhibits the case for the 2$^+_2$ state ($K$=2) against the 2$^+_1$ state ($K$=0) in $^{166}$Er.  Because of different $K$ values, the probability is down to 0.0008\% consistently with the arguments so far.  Thus, the present truncation works well to materialize the PT plot.

\end{document}